\documentclass[aps,prd,10pt,twocolumn,showpacs,showkeys,floatfix,superscriptaddress,nofootinbib]{revtex4-2}

\usepackage{amsmath,amssymb,amsfonts}
\usepackage{graphicx}
\usepackage{dcolumn}
\usepackage{bm}
\usepackage{silence}
\usepackage[
hypertexnames=false,
colorlinks=true,
linkcolor=blue,
citecolor=blue,
urlcolor=blue
]{hyperref}
\usepackage{xcolor}
\usepackage{booktabs}
\usepackage{multirow}
\usepackage{orcidlink}

\usepackage{tikz}
\usetikzlibrary{arrows.meta, positioning}
\begin{document}
	\raggedbottom
	\sloppy
	\hbadness=10000
	
	\title{Helical Inflationary Magnetogenesis
			through the Dark Photon Portal}
	
	\author{Debottam Nandi\, \orcidlink{0000-0002-7719-9623}}
	\affiliation{Department of Physics, School of Advanced Sciences, Vellore Institute of Technology (VIT) Chennai, Chennai 600127}

\begin{abstract}
	We investigate inflationary helical magnetogenesis
	through a kinetic portal connecting visible and
	dark Abelian gauge fields.
	An inflaton-driven Chern--Simons interaction
	amplifies the dark gauge field, while kinetic
	mixing controls its interaction with charged matter.
	For a given canonical dark-field amplification in the constant-mixing approximation, the visible magnetic power scales quadratically with the effective mixing, whereas Schwinger-induced dissipation scales cubically within the steady-state conductivity approximation.
	This distinct parametric dependence allows Schwinger dissipation to be suppressed more efficiently than visible magnetic power, without introducing a parametrically small gauge kinetic eigenvalue.
	We derive analytical consistency conditions and
	illustrate the framework through coupled gauge-field
	evolution in a Starobinsky inflationary
	background.
	Adopting unity as the nominal threshold for significant
	background backreaction and cumulative dissipation,
	we identify representative parameter choices yielding indicative present-day magnetic amplitudes
	of order $10^{-16}\,{\rm G}$ with sub-parsec wavelengths
	for instantaneous reheating.
	Current-free estimates approach $10^{-15}\,{\rm G}$
	near the onset of background backreaction, but exceed the adopted threshold for significant Schwinger dissipation.
	For an intervening matter-dominated reheating phase,
	the estimates scale as $T_{\rm reh}^{1/3}$.
	The surviving magnetic amplitudes depend on
	current-induced damping, magnetic survival through
	reheating and subsequent helical magnetohydrodynamic
	evolution. The framework also provides potential connections
	to dark radiation, baryogenesis and chiral
	gravitational waves.
\end{abstract}
	

	\maketitle
	
	%=========================================================

% Section I -- complete replacement for the Introduction.
% Assumes the numerical Starobinsky and astrophysical/cosmological sections
% have been incorporated, and the dedicated CMB section has been removed.
% Citation keys reconciled with main.bib.
\section{Introduction}
\label{sec:intro}

% 1. Observational motivation
Magnetic fields are ubiquitous in galaxies and galaxy clusters, yet their origin
remains an open problem in astrophysics and cosmology~\cite{Widrow:2002ud,
	Bernet:2008qp,Kronberg:2007dy,Grasso:2000wj,Kandus:2010nw,Giovannini:2003yn,
	Durrer:2013pga,Subramanian:2015lua,Vachaspati:2020blt}.
Although astrophysical batteries and dynamos can generate and amplify magnetic
fields, primordial seeds may contribute to the observed magnetization
~\cite{Subramanian:2019jyd,Seta:2020knq,Ryu:2011,Donnert:2018lbe}.
Blazar observations motivate investigations of intergalactic magnetic fields
~\cite{Neronov:2010,Tavecchio:2010,Taylor:2011,Finke:2015ona,HESS:2023zwb,
	MAGIC:2022piy,Burmeister:2025lgo}, while cosmic microwave background (CMB)
observations constrain primordial magnetic fields and their associated
perturbations~\cite{Caprini:2015gga,Planck:2015zrl,Trivedi:2013wqa}.

% 2. Conventional magnetogenesis
Primordial magnetogenesis has been investigated in early-Universe plasma
processes~\cite{Harrison:1970azx,Takahashi:2005nd,Fenu:2011}, cosmological phase
transitions~\cite{Hogan:1983zz,Quashnock:1988vs,Vachaspati:1991nm,Enqvist:1993np,
	Kibble:1995aa,Baym:1995fk,Sigl:1996dm}, and preheating
~\cite{Diaz-Gil:2007fch,Adshead:2015pva}.
Inflation can establish super-Hubble correlations, but the conformal invariance
of the free Maxwell action prevents vacuum amplification in a spatially flat
Friedmann background. Following the seminal works of
Refs.~\cite{Turner:1987bw,Ratra:1991bn}, numerous mechanisms involving curvature
couplings, evolving gauge kinetic functions and other non-minimal interactions
have been proposed~\cite{Dolgov:1993vg,Giovannini:2000ad,Giovannini:2001nh,
	Dimopoulos:2001wx,Bamba:2003av,Bamba:2004cu,Prokopec:2004au,Martin:2007ue}.
Subsequent developments include generalized gauge interactions, modified gravity
and effective-field-theory descriptions~\cite{Giovannini:2009xa,Giovannini:2013rza,
	Giovannini:2015hqa,Campanelli:2015jfa,Qian:2015mzl,Nandi:2017ajk,Kothari:2018aem,
	Talebian:2020drj,Giovannini:2021dso,Giovannini:2021due,Nandi:2021lpf,Maity:2021qps,
	Kushwaha:2022bwy,Tripathy:2022iev,Luongo:2026zfq}, as well as amplification during
departures from slow roll and reheating~\cite{Kobayashi:2014sga,Fujita:2016qab,
	Vilchinskii:2017qul,Kobayashi:2019uqs,Atkins:2025pvg,Ragavendra:2026fgs}.

% 3. Theoretical challenges
Despite these developments, three major challenges remain: the strong-coupling,
backreaction and Schwinger problems. An evolving Maxwell normalization can render
the effective electromagnetic coupling non-perturbative, while amplified gauge
fields can significantly modify the inflationary background or generate excessive
curvature perturbations~\cite{Demozzi:2009fu,Kanno:2009ei,Ferreira:2013sqa,
	Fujita:2014sna,Fujita:2012rb,Green:2015fss,Markkanen:2017kmy,Sharma:2017eps,
	Giovannini:2020zjo}. The reheating history further affects the attainable magnetic
amplitudes~\cite{Demozzi:2012wh,Haque:2020bip,Maity:2021qps,Tripathy:2021sfb,
	Dimopoulos:2024jnv}. Strong electric fields can also induce charged-particle
production and dissipative currents through the Schwinger mechanism
~\cite{Kobayashi:2014zza,Sobol:2018djj,Domcke:2018eki,Sharma:2017eps,
	Bastero-Gil:2025nfi}. Efficient magnetogenesis must therefore be reconciled with
perturbative control and limited gauge-field and charged-particle backreaction.

% 4. Axial magnetogenesis
A particularly attractive mechanism employs a parity-violating Chern--Simons
interaction that preferentially amplifies one gauge-field helicity
~\cite{Garretson:1992vt,Anber:2006xt,Durrer:2010mq,Anber:2009ua}.
This has motivated numerous realizations of helical magnetogenesis
~\cite{Caprini:2014mja,Fujita:2015iga,Adshead:2016iae,Sharma:2018kgs,
	Fujita:2019pmi,Shtanov:2019civ,Shtanov:2019gpx,Kushwaha:2020nfa,Talebian:2021dfq},
including connections to baryogenesis~\cite{Anber:2015yca,Domcke:2019mnd,
	Giovannini:2021xbi} and non-Abelian extensions~\cite{Brandenburg:2024awd}.
Although a purely axial interaction avoids the conventional strong-coupling
pathology associated with a vanishing Maxwell normalization, direct amplification
of the Standard Model hypercharge field remains susceptible to fermion production
and gauge-field backreaction~\cite{Sobol:2019xls,Gorbar:2021rlt,Gorbar:2021zlr,
	Fujita:2022fwc,vonEckardstein:2024tix}.

% 5. Recent progress and research question
Perturbative, gradient-expansion and lattice studies have advanced the understanding
of nonlinear axion--gauge-field dynamics~\cite{Cheng:2015oqa,Domcke:2020zez,
	Peloso:2022ovc,Caravano:2022epk,Figueroa:2023oxc,Domcke:2023tnn,Figueroa:2024rkr,Sobol:2026nfh}. Alam et al.~\cite{Alam:2024fid}
investigated the effects of gauge-field backreaction on natural inflation, while
Sharma et al.~\cite{Sharma:2024nfu} studied magnetic-field and gravitational-wave
production and emphasized the distinction between visible- and dark-sector gauge
amplification. Recent studies incorporating charged-particle currents highlight
both the importance of Schwinger screening and the possibility of post-inflationary
amplification~\cite{Iarygina:2025ncl,Sharma:2026bok}.
These developments motivate the central question of this
work: \emph{can helicity-selective gauge-field amplification
and its interaction with charged matter be controlled
separately, such that efficient visible magnetic-field
generation becomes possible without requiring a
parametrically small gauge kinetic normalization or
necessarily exposing the amplified gauge configuration
to strong charged-particle interactions throughout
inflation?}

% 6. Dark photons and earlier mechanisms
Kinetic mixing between visible and dark Abelian gauge fields provides a natural
setting for investigating this question~\cite{Holdom:1985ag,Abel:2008ai,
	Jaeckel:2010ni,Redondo:2010dp,Fabbrichesi:2020wbt}.
Previous studies have identified limitations of inflationary production in mixed
gauge sectors~\cite{Ferreira:2014mwa}, conductivity suppression of hidden-to-visible
magnetic transfer~\cite{Kamada:2018kyi}, and connections between axion-driven
magnetogenesis and dark radiation~\cite{Choi:2018dqr}.
Dark-photon-assisted inflationary magnetogenesis
with distinct gauge kinetic functions and transient
mixing has also been investigated~\cite{Nandi:2026xnw}, alongside related mechanisms involving ultralight dark photons, post-inflationary
axion magnetogenesis and dark-sector amplification during radiation domination
~\cite{Berger:2025pfv,Gerlach:2026sga,Alexander:2026hgt}.

% 7. Proposed mechanism
In contrast to the gauge-kinetic amplification
mechanism of Ref.~\cite{Nandi:2026xnw},
we investigate inflationary helical magnetogenesis
in which an inflaton-driven Chern--Simons interaction
amplifies a massless dark Abelian gauge field,
with no directly charged light matter, while a
kinetic portal generates a visible magnetic component. Unlike earlier
gauge-kinetic amplification mechanisms, the diagonal
Maxwell coefficients remain unity in the interaction
basis. In the constant-mixing approximation, at a
given canonical dark-field amplification, the visible
magnetic power scales quadratically with the effective
mixing, whereas Schwinger-induced dissipation scales
cubically within the steady-state conductivity
approximation. This distinct parametric dependence
allows Schwinger dissipation to be suppressed more
efficiently than visible magnetic power, without
introducing a parametrically small gauge kinetic
eigenvalue.

The constant-mixing analysis also motivates a dynamical
kinetic portal as a means of further enhancing the visible
magnetic field. For constant kinetic mixing, the visible
magnetic amplitude scales linearly with $|g_{\rm eff}|$,
and therefore remains suppressed for small mixing.
However, if the effective mixing evolves sufficiently
slowly from negligible values during most of inflation
to order unity towards its termination, the visible
magnetic field can acquire an additional amplification
relative to the constant small-mixing scenario.
Provided that the dynamically generated canonical visible
component remains subdominant, the visible magnetic
amplitude can then become comparable to that of the
amplified canonical dark field. Such a realization
combines suppressed charged-particle interactions during
most of inflation with enhanced visible magnetic-field
generation towards its termination. A subsequent
suppression of the kinetic mixing may render the visible
and dark sectors effectively decoupled at late times.

% 8. Scope and broader motivation
The generated magnetic helicity motivates the study of subsequent magnetic
correlation-scale growth in a conducting plasma~\cite{Brandenburg:1996fc,
	Christensson:2000sp,Banerjee:2004df,Brandenburg:2016odr,Brandenburg:2017neh,
	Hosking:2020wom,Zhou:2022xhk,Hosking:2022umv,Brandenburg:2024tyi}.
Beyond magnetic seeds, the surviving dark component, hypermagnetic helicity and
parity-violating gauge stresses may connect the mechanism to dark radiation,
baryogenesis and chiral gravitational waves~\cite{Anber:2015yca,Domcke:2019mnd,
	Kamada:2018kyi,Hamada:2025ooy,Sorbo:2011rz,Adshead:2019igv}.
Although these connections have been explored in
various settings~\cite{Adshead:2016iae,Gorbar:2021rlt,
	Sharma:2024nfu,Choi:2018dqr,Ragavendra:2026fgs},
the present framework provides a common physical
origin through a helically amplified, kinetically
mixed inflationary gauge sector. We complement the constant-mixing analytical treatment
	with the numerical evolution of the coupled gauge modes
	in Starobinsky inflation~\cite{Starobinsky:1980te,
	Planck:2018jri}, considering a representative phenomenological
	realization of the evolving kinetic portal. We
	subsequently estimate the present-day magnetic
	amplitude, its characteristic correlation scale,
	and the associated dark radiation abundance under
	prescribed post-inflationary evolution. Our primary
	objective is to establish the analytical properties
	and magnetic-generation potential of the framework.
	A self-consistent treatment of the induced currents,
	the detailed reheating dynamics, and sourced
	cosmological perturbations~\cite{Barnaby:2010vf,Barnaby:2011qe} is beyond the scope of this work.

% 9. Organization and conventions
Section \ref{sec:setup} introduces the model and canonical gauge
basis, while Sec.~\ref{sec:gauge_production} examines inflationary gauge
production, Schwinger dissipation, and background
consistency conditions in the constant-mixing
approximation. Section \ref{sec:magnetic_end} investigates magnetic-field
generation and extends the analytical framework to
a broader class of phenomenological kinetic portals.
Section \ref{sec:postinflation} describes the subsequent magnetic-field
evolution. In Sec.~\ref{sec:numerical}, we illustrate the framework
through the numerical evolution of the coupled gauge
fields in Starobinsky inflation. Section \ref{sec:implications} discusses
the associated astrophysical and cosmological
implications, and Sec.~\ref{sec:conclusions} concludes.

We use natural units $c=\hbar=k_{\rm B}=1$, the reduced Planck mass
$M_{\rm Pl}=(8\pi G)^{-1/2}$, and metric signature $(-,+,+,+)$.
Primes and dots denote conformal- and cosmic-time derivatives, respectively.	
	%=========================================================
	\section{Model and Canonical Gauge Basis}
	\label{sec:setup}
	
	\subsection{Action and Coulomb gauge}
	\label{subsec:action-coulomb-gauge}
	
	We consider a canonical pseudoscalar $\phi$ coupled to two massless
	$U(1)$ gauge fields on a spatially flat FLRW background,
	$ds^2=a^2(\tau)(-d\tau^2+d\mathbf{x}^2)$. 
	The action (with the matter sector represented by its current) is
	\begin{align}
		S=\int d^4x\sqrt{-g}\,\Big[&\frac{M_{\rm Pl}^2}{2}R
		-\frac12\partial_\mu\phi\partial^\mu\phi-V(\phi)
		\nonumber\\
		&-\frac14 F_{\mu\nu}F^{\mu\nu}-\frac14 Y_{\mu\nu}Y^{\mu\nu}
		\nonumber\\
		&-\frac{\epsilon_{\rm mix}}2F_{\mu\nu}Y^{\mu\nu}
		+\frac{\alpha\phi}{4f}\tilde Y^{\mu\nu}Y_{\mu\nu}
		+A_\mu J^\mu\Big].
		\label{eq:action}
	\end{align}
	Here $F_{\mu\nu}=\partial_\mu A_\nu-\partial_\nu A_\mu$ and
	$Y_{\mu\nu}=\partial_\mu X_\nu-\partial_\nu X_\mu$. In general, the kinetic-mixing parameter may be a
	function of the scalar field,
	$\epsilon_{\rm mix}=\epsilon_{\rm mix}(\phi)$.
	For the analytical treatment presented in the
	following sections, we first consider the
	constant-mixing limit. This allows us to isolate
	the essential physical properties of the kinetic
	portal independently of its detailed evolution.
	We subsequently extend the framework to
	phenomenological realizations with evolving
	kinetic mixing. The interaction-basis
	field $X_\mu$ has no direct Standard Model charge, whereas $A_\mu$ couples
	to the current $J^\mu \equiv e j^\mu$. Above the electroweak scale, the visible Abelian field
	is identified with the Standard Model hypercharge field.
	We neglect a separate electroweak transfer factor in
	our magnetic-field estimates. We define the dual field strength by
	\begin{equation}
		\tilde{Y}^{\mu\nu} \equiv \frac{1}{2\sqrt{-g}} \varepsilon^{\mu\nu\rho\sigma} Y_{\rho\sigma} , \quad \varepsilon^{0123} = +1 ,
	\end{equation}
	where $\varepsilon^{\mu\nu\rho\sigma}$ is the totally antisymmetric Levi--Civita symbol. We impose Coulomb gauge, $\partial_i A_i=\partial_i X_i=0$.
	For the homogeneous background $\phi=\phi(\tau)$, the temporal
	components $A_0$ and $X_0$ are non-dynamical and satisfy constraint
	equations. In the absence of charge density and with vanishing
	longitudinal boundary modes, they can consistently be set to zero. After
	eliminating the non-dynamical and longitudinal degrees of
	freedom and integrating by parts, the quadratic Lagrangian
	for the transverse spatial modes becomes
	\begin{eqnarray} \label{eq:Lagrangian_Coulomb}
		\mathcal{L} &=& \frac{1}{2}A_i'^2 - \frac{1}{2}(\partial_i A_j)^2 + \frac{1}{2}X_i'^2 - \frac{1}{2}(\partial_i X_j)^2 \nonumber \\
		&&+ \epsilon_{\rm mix} \left( A_i'X_i' - (\partial_i A_j)(\partial_i X_j)\right) \nonumber \\
		&&- \frac{\alpha\phi'}{2f}\varepsilon^{ijk}X_i\partial_j X_k + a^2 J_i A_i ,
	\end{eqnarray}
	where primes denote $\partial_\tau$, $\varepsilon^{ijk}$ is the spatial
	Levi--Civita symbol, and $J_i$ denotes the transverse part of the covariant
	spatial current. Spatial indices on the conformal fields are contracted
	with $\delta_{ij}$.
	
	\subsection{Canonical gauge basis}
	\label{subsec:canonical-gauge-basis}
	For a time-independent kinetic-mixing parameter $\epsilon_{\rm mix}$, the kinetic
	mixing is removed by the canonical transformation
	
	\begin{equation} \label{eq:canonical_transform}
		A_i =  \tilde{A}_i - g_{\text{eff}} \tilde{X}_i , \quad X_i = \frac{\tilde{X}_i}{\beta} ,
	\end{equation}
	where the normalization and effective coupling are
	\begin{equation}\label{eq:geff_constant}
		\beta \equiv \sqrt{1 - \epsilon_{\rm mix}^2}  , \quad g_{\text{eff}} \equiv \frac{\epsilon_{\rm mix}}{\beta}, \quad |\epsilon_{\rm mix}| < 1.
	\end{equation}
	Equivalently, the original kinetic-mixing parameter
	and the canonical normalization can be expressed in
	terms of the effective mixing as
	$\epsilon_{\rm mix}=g_{\rm eff}/\sqrt{1+g_{\rm eff}^2}$
	and $\beta^2=(1+g_{\rm eff}^2)^{-1}$, respectively. Substitution into Eq.~\eqref{eq:Lagrangian_Coulomb} gives
	\begin{eqnarray} \label{eq:Lagrangian_canonical}
		\mathcal{L} &=& \frac{1}{2}\tilde{A}_i'^2 - \frac{1}{2}(\partial_j \tilde{A}_i)^2 + \frac{1}{2}\tilde{X}_i'^2 - \frac{1}{2}(\partial_j \tilde{X}_i)^2 \nonumber \\
		&&- \frac{\alpha\phi'}{2f\beta^2}\varepsilon^{ijk}\tilde{X}_i\partial_j \tilde{X}_k + a^2 J_i \tilde{A}_i - a^2 g_{\text{eff}} J_i \tilde{X}_i .
	\end{eqnarray}
	The interaction terms in Eq.~(\ref{eq:Lagrangian_canonical}) show that visible
	charged matter couples to the canonical visible and
	dark gauge fields with couplings $e$ and
	$-eg_{\rm eff}$, respectively. For
	$|g_{\rm eff}|\lesssim 1$, both interactions remain
	perturbative for perturbative $e$. Moreover, the
	eigenvalues of the original gauge kinetic matrix,
	$\kappa_\pm=1\pm\epsilon_{\rm mix}$, remain positive
	and of order unity in this regime. The framework
	therefore avoids the conventional gauge-kinetic
	strong-coupling problem.
	
	For constant kinetic mixing, the canonical
	transformation generates no additional derivative
	couplings. In the absence of charged currents, the
	two canonical gauge fields dynamically decouple,
	and only $\widetilde{X}_i$ experiences the axial
	instability. This limit constitutes the basis of
	our analytical investigation. For evolving kinetic
	mixing, the transformation continues to diagonalize
	the instantaneous kinetic matrix, but additional
	derivative couplings arise and the two canonical
	fields no longer evolve independently. We return to
	this more general situation when investigating
	phenomenological realizations of the kinetic portal. Defining
	\begin{align}
		\mathcal F_{\mu\nu}&=\partial_\mu\widetilde A_\nu-\partial_\nu\widetilde A_\mu,\\
		\mathcal Y_{\mu\nu}&=\partial_\mu\widetilde X_\nu-\partial_\nu\widetilde X_\mu,
	\end{align}
	we obtain, for constant kinetic mixing,
	\begin{equation}
		F_{\mu\nu}=\mathcal F_{\mu\nu}-g_{\rm eff}\mathcal Y_{\mu\nu},\qquad
		Y_{\mu\nu}=\mathcal Y_{\mu\nu}/\beta.
	\end{equation}
	\subsection{Schwinger effects in the kinetically mixed model}
	\label{subsec:CircumSchwinger}
	We include no light particles charged directly under the
	dark $U(1)$. Equation~\eqref{eq:Lagrangian_canonical}
	implies an effective charge $|e_{\rm eff}|=e|g_{\rm eff}|$
	for Standard Model matter coupled to the canonical dark
	mode. For constant kinetic mixing, when the canonical
	dark field is amplified while the canonical visible
	field remains unamplified, we obtain
	\begin{equation}\label{eq:vis-can-amp}
		\mathbf E_A^{(\rm amp)}\simeq-g_{\rm eff}\mathbf E_{\widetilde X},
		\qquad
		\mathbf B_A^{(\rm amp)}\simeq-g_{\rm eff}\mathbf B_{\widetilde X}.
	\end{equation}
	Unlike conventional axial magnetogenesis, where the
	amplified gauge field couples directly to charged matter,
	the interaction of the amplified dark field with visible
	charged particles is controlled by kinetic mixing.
	For $|\epsilon_{\rm mix}|\ll 1$, one has
	$|g_{\rm eff}|\simeq|\epsilon_{\rm mix}|\ll 1$ and hence
	$|e_{\rm eff}|\ll e$, suppressing Schwinger-induced
	dissipation at a given canonical dark-field amplitude.
	The kinetic portal thus provides an additional degree
	of freedom in controlling Schwinger screening, which
	we quantify in Sec.~\ref{sec:gauge_production}.

	% Replacement for Section III of helical manuscript.
	% Requires amsmath, amssymb; BibTeX keys in section_III_references.bib.
	% Existing Section IV equation numbering will change unless explicitly fixed.
	\section{Inflationary gauge production and consistency conditions}
	\label{sec:gauge_production}
	
For constant kinetic mixing and negligible charged
currents, the canonical visible field remains free,
while the axial interaction amplifies one helicity of
the canonical dark field. The effective mixing controls
both the instability parameter and the projection onto
the current-coupled visible sector. We decompose the
transverse canonical fields into helicity eigenstates~
\cite{Kamada:2018kyi, Ferreira:2014mwa}:
	\begin{align}
		\tilde A_i(\tau,\mathbf{x})
		&=\sum_{\lambda=\pm}\int\frac{d^3k}{(2\pi)^{3/2}}
		\hat a_\lambda(\mathbf{k})\,\tilde A_\lambda(k,\tau)\nonumber\\
		&\quad\times e^{\lambda}_{i}(\hat{\mathbf{k}})e^{i\mathbf{k}\cdot\mathbf{x}}
		+\text{h.c.},\\
		\tilde X_i(\tau,\mathbf{x})
		&=\sum_{\lambda=\pm}\int\frac{d^3k}{(2\pi)^{3/2}}
		\hat x_\lambda(\mathbf{k})\,\tilde X_\lambda(k,\tau)\nonumber\\
		&\quad\times e^{\lambda}_{i}(\hat{\mathbf{k}})e^{i\mathbf{k}\cdot\mathbf{x}}
		+\text{h.c.}
		\label{eq:helicity-expansions}
	\end{align}
	with commutation relations 
	\begin{align}
		[\hat{a}_\lambda(\bm{k}),\hat{a}^\dagger_{\lambda'}(\bm{k}')]
		&= \delta_{\lambda\lambda'}\delta^{(3)}(\mathbf{k}-\mathbf{k}'), \nonumber \\
		[\hat{x}_\lambda(\bm{k}),\hat{x}^\dagger_{\lambda'}(\bm{k}')] &= \delta_{\lambda\lambda'}\delta^{(3)}(\mathbf{k}-\mathbf{k}'), \nonumber \\
		[\hat{x}_\lambda(\bm{k}),\hat{a}^\dagger_{\lambda'}(\bm{k}')] &= [\hat{a}_\lambda(\bm{k}),\hat{x}^\dagger_{\lambda'}(\bm{k}')] = 0. \nonumber
	\end{align}
	The polarization vectors satisfy $k_i e^\lambda_i(\hat{\bm{k}})=0,$ $e_i^\lambda (\hat{\bm k}) e_i^{\lambda' *}(\hat{\bm{k}}) = \delta_{\lambda \lambda'}$ and
	$i\varepsilon_{ijk}k_j e^\lambda_k(\hat{\bm{k}})=\lambda k\,e^\lambda_i(\hat{\bm{k}})$. In the quasi-de Sitter approximation, the resulting
	mode equation takes the standard axion--gauge form,
	with the instability parameter re-scaled by the
	kinetic normalization \cite{Anber:2006xt, Anber:2009ua, Barnaby:2011qe}:
	\begin{align}
				\label{eq:III_mode}
		&\widetilde X_\lambda''+\left(k^2+\frac{2\lambda k\xi_{\rm eff}}{\tau}\right)\widetilde X_\lambda=0,\\
		&\xi_{\rm eff}\equiv-\frac{\alpha\dot\phi}{2fH\beta^2},
		\qquad \beta^2=1-\epsilon_{\rm mix}^2 . \nonumber
	\end{align}
	Even for constant mixing, $\xi_{\rm eff}$ generally varies. For a canonical
	scalar on an un-backreacted background, $\epsilon_H = -\dot H/H^2$ gives
	\begin{equation}
		|\xi_{\rm eff}|=\frac{|\alpha|M_{\rm Pl}}{f\beta^2}
		\sqrt{\frac{\epsilon_H}{2}}.
		\label{eq:xi_slow_roll}
	\end{equation}
	For constant kinetic mixing, $\xi_{\rm eff}$ remains
	constant if $\epsilon_H$ is assumed to be constant
	throughout inflation. For the analytical treatment,
	we adopt this assumption and work in the quasi-de Sitter
	approximation, $a\simeq-1/(H_{\rm inf}\tau)$. For $\xi_{\rm eff}>0$, the $+$ helicity is tachyonic in $0<x\equiv-k\tau<2\xi_{\rm eff}$. With adiabatic vacuum normalization, the amplified
	positive-helicity mode is approximately given by
	the growing WKB solution~\cite{Anber:2006xt,Anber:2009ua}:
	\begin{align}
		\widetilde X_+(k,\tau)&\simeq\frac{e^{\pi\xi_{\rm eff}-2\sqrt{2\xi_{\rm eff}x}}}{\sqrt{2k}}
		\left(\frac{x}{2\xi_{\rm eff}}\right)^{1/4}.
		\label{eq:III_wkb}
	\end{align}
This approximation is valid for large $\xi_{\rm eff}$
over the interval
$1/(8\xi_{\rm eff})\lesssim x\ll2\xi_{\rm eff}$.
We use it to obtain analytical estimates of the
gauge-field energy, backreaction and magnetic-field
generation.

	The canonical electric and magnetic components in the local orthonormal
	frame (with spatial Euclidean indices) are
	\begin{align}
		E_{\tilde{A}}^i&=-\frac{1}{a^2}\tilde{A}_i',
		\qquad
		B_{\tilde{A}}^i=\frac{1}{a^2}\varepsilon^{ijk}\partial_j \tilde{A}_k , \\
		E_{\tilde{X}}^i&=-\frac{1}{a^2}\tilde{X}_i',
		\qquad
		B_{\tilde{X}}^i=\frac{1}{a^2}\varepsilon^{ijk}\partial_j \tilde{X}_k ,
		\label{eq:can-field-definitions}
	\end{align}
	
	For constant mixing, the interaction-basis visible
	field is
	$A_i=\widetilde A_i-g_{\rm eff}\widetilde X_i$.
	Since the canonical oscillators are initially
	uncorrelated, the physical visible magnetic power
	spectrum is
	\begin{equation}
		\mathcal P_{B_A}(k,\tau)
		\equiv
		\frac{d\langle B_A^2\rangle}{d\ln k}
		=
		\frac{k^5}{2\pi^2a^4}
		\sum_{\lambda=\pm}
		\left[
		|\widetilde A_\lambda|^2
		+
		g_{\rm eff}^2|\widetilde X_\lambda|^2
		\right].
		\label{eq:III_visible_power}
	\end{equation}
	The spectral electric and magnetic energy densities
of the canonical gauge fields are
\begin{align}
	\frac{d\rho_{\widetilde V}^{(E)}}{d\ln k}
	&=
	\frac{k^3}{4\pi^2a^4}
	\sum_{\lambda=\pm}
	\left|\widetilde V'_\lambda(k,\tau)\right|^2,
	\nonumber\\
	\frac{d\rho_{\widetilde V}^{(B)}}{d\ln k}
	&=
	\frac{k^5}{4\pi^2a^4}
	\sum_{\lambda=\pm}
	\left|\widetilde V_\lambda(k,\tau)\right|^2,
	\label{eq:drhoFdlnk}
\end{align}
where $\widetilde V=\widetilde A,\widetilde X$.
The gauge-sector energy density is most
conveniently defined in the canonical basis,
\begin{equation}
	\rho_g=
	\frac12
	\left\langle
	\widetilde{\mathbf E}_A^2+
	\widetilde{\mathbf B}_A^2+
	\widetilde{\mathbf E}_X^2+
	\widetilde{\mathbf B}_X^2
	\right\rangle.
	\label{eq:III_totalenergy}
\end{equation}
Throughout the analytical treatment, the spectral
integrals are restricted to the tachyonic production
band. For sufficiently large $\xi_{\rm eff}$, the
gauge-field energy is dominated by the exponentially
amplified canonical dark component. Consequently,
\begin{equation}
	\rho_g
	\simeq
	\rho_{\widetilde X}^{(E)}
	+
	\rho_{\widetilde X}^{(B)}.
	\label{eq:III_produced_energy}
\end{equation}

The axial Chern--Simons interaction does not
contribute directly to the metric stress tensor
when its coupling is treated as a prescribed
scalar function.
	The amplified gauge field also sources the
	homogeneous inflaton equation,
	\begin{equation}
		\ddot\phi+3H\dot\phi+V_{,\phi}
		=
		\frac{\alpha}{4f\beta^2}
		\left\langle
		\mathcal{Y}_{\mu\nu}
		\widetilde{\mathcal{Y}}^{\mu\nu}
		\right\rangle
		=
		-\frac{\alpha}{f\beta^2}
		\left\langle
		\widetilde{\mathbf E}_X\cdot
		\widetilde{\mathbf B}_X
		\right\rangle .
		\label{eq:inflaton-backreaction-eom}
	\end{equation}
	Defining $u_\lambda(x)=\sqrt{2k}\,
	\widetilde X_\lambda(k,\tau)$, the magnetic and
	electric energy densities integrated over the
	tachyonic production band, $0<x<2\xi_{\rm eff}$,
	are given by
	\begin{equation}
		\rho_{\widetilde X}^{(B)}
		=
		\frac{H_{\rm inf}^4}{8\pi^2}
		\sum_{\lambda=\pm1}
		\int_0^{2\xi_{\rm eff}}
		dx\,x^3|u_\lambda(x)|^2,
		\label{eq:rho_B_integral}
	\end{equation}
	and
	\begin{equation}
		\rho_{\widetilde X}^{(E)}
		=
		\frac{H_{\rm inf}^4}{8\pi^2}
		\sum_{\lambda=\pm1}
		\int_0^{2\xi_{\rm eff}}
		dx\,x^3
		\left|\frac{du_\lambda}{dx}\right|^2
		.
		\label{eq:rho_E_integral}
	\end{equation}
	The corresponding axial source is
	\begin{equation}
		\left\langle
		\mathbf E_{\widetilde X}\cdot
		\mathbf B_{\widetilde X}
		\right\rangle
		=
		\frac{H_{\rm inf}^4}{8\pi^2}
		\sum_{\lambda=\pm1}\lambda
		\int_0^{2\xi_{\rm eff}}dx\,x^3
		\frac{d|u_\lambda|^2}{dx}.
		\label{eq:EB_integral}
	\end{equation}
	 Using the growing WKB solution, retaining only the
	 leading exponentially enhanced contributions, 
	 we obtain~\cite{Anber:2006xt,Anber:2009ua,
		Barnaby:2011qe}:
	\begin{align}
		&\rho_{\widetilde X}^{(E)}\simeq1.39\times10^{-4}H_{\rm inf}^4
		\frac{e^{2\pi\xi_{\rm eff}}}{\xi_{\rm eff}^3}, \\ 
		&\rho_{\widetilde X}^{(B)}\simeq1.22\times10^{-4}H_{\rm inf}^4
		\frac{e^{2\pi\xi_{\rm eff}}}{\xi_{\rm eff}^5},\label{eq:III_energies}\\
		&\bigl|\langle\widetilde{\mathbf E}_X\cdot\widetilde{\mathbf B}_X\rangle\bigr|
		\simeq2.44\times10^{-4}H_{\rm inf}^4
		\frac{e^{2\pi\xi_{\rm eff}}}{\xi_{\rm eff}^4}.
		\label{eq:III_EB}
	\end{align}
	We quantify background consistency using the
	produced gauge-energy fraction, the axial
	source relative to Hubble friction, and the
	gauge contribution to the Hubble slow-roll
	parameter:
	\begin{align}
		\delta_g
		&\equiv
		\frac{\rho_g}
		{3M_{\rm Pl}^2H^2},
		\label{eq:III_dg}\\
		R_\phi
		&\equiv
		\frac{|\alpha|}
		{3f\beta^2H|\dot\phi|}
		\left|
		\left\langle
		\widetilde{\mathbf E}_X\cdot
		\widetilde{\mathbf B}_X
		\right\rangle
		\right|,
		\label{eq:III_Rphi}\\
		R_H
		&\equiv
		\frac{2\delta_g}{\epsilon_H^{(0)}}.
		\label{eq:III_RH}
	\end{align}
	where $\epsilon_H^{(0)}\equiv
	\dot{\phi}_0^2/(2M_{\rm Pl}^2H_0^2)$
	is evaluated on the prescribed scalar-only
	background. For an approximately isotropic gauge-field configuration,
	$p_g=\rho_g/3$, the second Friedmann equation gives
	\begin{equation}
		-\frac{\dot H}{H^2}
		=\frac{\dot\phi^2}{2M_{\rm Pl}^2H^2}+2\delta_g.
	\end{equation}
	Thus, $R_H$ measures the gauge-field contribution to
	the Hubble slow-roll parameter relative to the
	scalar-only background contribution. Negligible background backreaction requires
	$\delta_g,R_\phi,R_H\ll1$.
	Throughout this work, unity is adopted as the nominal
	threshold for the onset of significant backreaction.
	Our objective is to estimate this threshold on a
	prescribed inflationary background rather than to
	investigate the subsequent backreacted evolution.
	
	In the following analytical estimates,
	$H$ and $\epsilon_H$ denote their
	prescribed scalar-only background values,
	$H_0$ and $\epsilon_H^{(0)}$, respectively.
	In particular,
	$\dot\phi_0^2=
	2\epsilon_H^{(0)}M_{\rm Pl}^2H_0^2$.
	Using Eq.~\eqref{eq:III_EB}, we obtain
	\begin{equation}
		R_\phi\simeq
		8.12\times10^{-5}
		\frac{H_{\rm inf}^2}
		{\epsilon_HM_{\rm Pl}^2}
		\frac{e^{2\pi\xi_{\rm eff}}}
		{\xi_{\rm eff}^3}.
		\label{eq:III_Rphi_wkb}
	\end{equation}
	These conditions diagnose the validity of
	the un-backreacted background approximation
	rather than providing observational bounds.
	Gauge-sourced scalar and tensor perturbations
	require separate consideration
	\cite{Barnaby:2011qe}.
	
For illustration, assuming that the scalar power
spectrum is dominated by vacuum fluctuations,
we use the leading Starobinsky slow-roll
approximation at $N=60$ with the measured scalar
amplitude $A_s\simeq2.1\times10^{-9}$
\cite{Planck:2018vyg}:
\begin{align}
	\epsilon_{H,60}
	&\simeq\frac{3}{4N^2}
	=2.083\times10^{-4},
	\nonumber\\
	H_{\rm inf}\equiv H_*
	&=
	M_{\rm Pl}
	\sqrt{8\pi^2\epsilon_{H,60}A_s}
	\simeq5.88\times10^{-6}M_{\rm Pl}.
	\label{eq:H_COBE}
\end{align}
Here $H_*$ denotes the pivot-scale Hubble
parameter. The constant-$H_{\rm inf}$
approximation is used only as an analytical
reference and does not represent the
dynamically determined Starobinsky endpoint.

Saturating each of the background backreaction
criteria at unity, the WKB estimates give
\begin{equation}
	\begin{aligned}
		\xi_{\rm eff}^{(g)}&\simeq 6.30,\\
		\xi_{\rm eff}^{(\phi)}&\simeq 4.73,\\
		\xi_{\rm eff}^{(H)}&\simeq 4.70.
	\end{aligned}
	\label{eq:III_thresholds}
\end{equation}
We emphasize that
these values diagnose the onset of significant
backreaction and are not bounds ensuring
perturbative control. Combining the three conditions, we obtain the
upper limit
\begin{equation}
	\max\{\delta_g,R_\phi,R_H\}\lesssim 1
	\quad\Rightarrow\quad
	\xi_{\rm eff}\lesssim 4.70.
	\label{eq:xi_eff_max}
\end{equation}
For comparison, imposing the stronger requirement
$\max\{\delta_g,R_\phi,R_H\}\lesssim0.01$ gives
$\xi_{\rm eff}\lesssim3.87$ for the same reference
background.  This constant-instability reference is not intended
to represent the same axial coupling as the evolving
Starobinsky benchmarks considered in Sec.~\ref{sec:numerical}. At the backreaction threshold $\xi_{\rm eff}\simeq4.70$, it gives
\begin{align}
	\rho_{\widetilde X}^{(B)}
	&\simeq 3.55\times10^5H_{\rm inf}^4,\\
	\rho_{\widetilde X}^{(E)}
	&\simeq8.96\times10^6H_{\rm inf}^4.
	\label{eq:rho_E_numerical}
\end{align}
The associated gauge-energy fraction is
$\delta_g\simeq1.07\times10^{-4}$, while
$R_H\simeq1, ~R_\phi \simeq 0.87$. Thus, despite the small
gauge-energy fraction, the gauge-field contributions
to the inflaton and Hubble evolution become
comparable to their scalar-only counterparts.
This benchmark therefore represents the onset of
significant background backreaction.

For Standard Model hypercharge, the Schwinger-induced
conductivity in the strong-field, approximately
collinear limit can be estimated as
\begin{equation}
	\sigma_{\rm Sch}\simeq
	\frac{41g'^3}{72\pi^2H}
	B_A\coth\left(\frac{\pi B_A}{E_A}\right),
	\label{eq:schwinger_conductivity}
\end{equation}
Here $E_A$ and $B_A$ denote the RMS amplitudes of
the amplified visible electric and magnetic fields,
$g'$ denotes the hypercharge coupling, and the
contributing fermions are assumed to be effectively
massless
\cite{Domcke:2018eki,Gorbar:2021rlt,Gorbar:2021zlr}.
Equation~\eqref{eq:schwinger_conductivity} provides
a steady-state conductivity estimate in the
strong-field regime, $g'E_A/H^2\gtrsim1$, rather
than an exact local current for the stochastic
gauge-field configuration.

For constant kinetic mixing and given canonical
dark-field spectra, Eq.~(\ref{eq:vis-can-amp}) implies the following
scaling relations within the stationary strong-field
approximation:
\begin{equation}
	\sigma_{\rm Sch}\propto|g_{\rm eff}|,\qquad
	\mathcal P_{\rm Sch}
	=\sigma_{\rm Sch}E_A^2
	\propto|g_{\rm eff}|^3.
\end{equation}
The visible magnetic power scales quadratically
with the effective mixing, whereas the stationary
estimate of Schwinger-induced dissipation scales
cubically at a given canonical dark-field spectrum.
Consequently, the dissipated power relative to
the visible magnetic energy per Hubble time,
\begin{equation}
	\frac{\mathcal P_{\rm Sch}}{H\rho_{B_A}}
	=
	2\frac{\sigma_{\rm Sch}}{H}
	\frac{E_A^2}{B_A^2}
	\propto |g_{\rm eff}|,
\end{equation}
exhibits a linear suppression.
This advantage is accompanied by a trade-off:
at fixed visible magnetic amplitude, the required
canonical dark magnetic energy scales as
$|g_{\rm eff}|^{-2}$.
The kinetic portal therefore provides additional
freedom in balancing charged-particle dissipation
against gauge-field backreaction.

For the nominal analytical benchmark
$\xi_{\rm eff}\simeq4.70$, adopting $g'=0.35$
gives the stationary conductivity estimate
\begin{equation}
	\frac{\sigma_{\rm Sch}}{H}
	\simeq3.76|g_{\rm eff}|.
\end{equation}
For the conservative reference
$\xi_{\rm eff}\simeq3.87$, the corresponding WKB
estimate is approximately $0.39|g_{\rm eff}|$.

To characterize the instantaneous importance of
Schwinger dissipation relative to the produced
gauge-sector energy, we define
\begin{equation}
	\mathcal D_{\rm Sch}
	\equiv
	\frac{\sigma_{\rm Sch}E_A^2}
	{H\rho_g}
	=
	2g_{\rm eff}^2\frac{\sigma_{\rm Sch}}{H}
	\frac{\rho_{\widetilde X}^{(E)}}
	{\rho_g}.
	\label{eq:schwinger_dissipation}
\end{equation}
For $\xi_{\rm eff}=3.87$ and $4.70$, this gives
$\mathcal D_{\rm Sch}\simeq0.74|g_{\rm eff}|^3$
and $7.23|g_{\rm eff}|^3$, respectively. We adopt $\mathcal D_{\rm Sch}\sim1$ as the nominal
threshold for appreciable instantaneous dissipation
relative to the total produced gauge energy over a
Hubble time. At the nominal backreaction benchmark
$\xi_{\rm eff}\simeq4.70$, this corresponds to
$|g_{\rm eff}|\simeq0.52$. The cumulative importance
of dissipation, including its normalization to the
visible gauge energy, is investigated in Sec.~\ref{sec:numerical}. 

These estimates motivate phenomenological realizations
in which the effective mixing remains suppressed during
most of inflation and approaches
$|g_{\rm eff,end}|\sim\mathcal{O}(1)$ towards its
termination. Although such a prescription may reduce
cumulative Schwinger dissipation while allowing efficient
visible magnetic production, its viability depends on
the coupled gauge-field evolution and the charged-particle
response. We explore this possibility in the following
sections.

\section{Magnetic Field at the End of Inflation}
\label{sec:magnetic_end}

Following the standard treatment of helical inflationary
magnetogenesis~\cite{Anber:2006xt,Anber:2009ua,
	Fujita:2015iga,Adshead:2016iae},
we characterize the generated magnetic field by its
root-mean-square (RMS) amplitude and magnetic integral
scale. In the constant-mixing approximation, the
visible magnetic spectrum associated with the dark
gauge modes is
\begin{equation}
	\mathcal P_{B_A}^{\rm amp}(k,\tau)
	\simeq
	\frac{g_{\rm eff}^2 k^4}{4\pi^2a^4}
	\sum_{\lambda=\pm}|u_\lambda(x)|^2,
	\label{eq:BA_amplified}
\end{equation}
Substituting the growing WKB solution of Sec.~III
and retaining the dominant positive-helicity mode,
the visible magnetic spectrum becomes
\begin{equation}
	\mathcal P_{B_A}^{\rm amp}(k,\tau)
	\simeq
	\frac{g_{\rm eff}^2H_{\rm inf}^4}{4\pi^2}
	\frac{x^{9/2}}{\sqrt{2\xi_{\rm eff}}}
	e^{2\pi\xi_{\rm eff}-4\sqrt{2\xi_{\rm eff}x}}.
\end{equation}
The WKB spectrum is blue below its peak and
exponentially suppressed at sufficiently large
momenta within the validity of the approximation. However, in the strict infrared limit (large scales), where the growing
WKB approximation ceases to be valid, the
constant-$\xi_{\rm eff}$ mode equation yields
$\mathcal P_{B_A}^{\rm amp}\propto k^4$.

The physical
RMS magnetic field at the end of inflation is
\begin{align}
	B_{\rm end}^2
	&\equiv
	\int_{\rm amp}d\ln k\,
	\mathcal P_{B_A}^{\rm amp}(k,\tau_{\rm end})
	\nonumber\\
	&\simeq
	\frac{g_{\rm eff}^2 H_{\rm inf}^4}{4\pi^2}
	\sum_{\lambda=\pm}
	\int_0^{2\xi_{\rm eff}}dx\,
	x^3|u_\lambda(x)|^2.
	\label{eq:Bend_rms}
\end{align}
Using the growing WKB solution of Sec.~\ref{sec:gauge_production},
retaining the leading exponentially enhanced
contribution and extending the integration limits
to $0<x<\infty$, we obtain
\begin{equation}
	B_{\rm end}
	\simeq
	1.56\times10^{-2}|g_{\rm eff}|H_{\rm inf}^2
	\frac{e^{\pi\xi_{\rm eff}}}
	{\xi_{\rm eff}^{5/2}}.
	\label{eq:Bend_analytic}
\end{equation}
The characteristic physical magnetic integral scale
is defined as~\cite{Anber:2006xt}
\begin{equation}
	L_{c,\rm end}^{\rm phys}
	\equiv
	a_{\rm end}
	\frac{
		\displaystyle\int_{\rm amp}d\ln k\,
		k^{-1}\mathcal P_{B_A}^{\rm amp}(k,\tau_{\rm end})
	}{
		\displaystyle\int_{\rm amp}d\ln k\,
		\mathcal P_{B_A}^{\rm amp}(k,\tau_{\rm end})
	}.
	\label{eq:Lc_end}
\end{equation}
In terms of the dimensionless momentum variable $x$,
this expression becomes
\begin{equation}
	H_{\rm inf}L_{c,\rm end}^{\rm phys}
	\simeq
	\frac{
		\displaystyle\sum_{\lambda=\pm}
		\int_0^{2\xi_{\rm eff}}dx\,
		x^2|u_\lambda(x)|^2
	}{
		\displaystyle\sum_{\lambda=\pm}
		\int_0^{2\xi_{\rm eff}}dx\,
		x^3|u_\lambda(x)|^2
	}.
	\label{eq:Lc_end_integral}
\end{equation}
The leading WKB approximation yields
\begin{equation}
	L_{c,\rm end}^{\rm phys}
	\simeq
	\frac{4\xi_{\rm eff}}{7H_{\rm inf}},
	\label{eq:Lc_end_analytic}
\end{equation}
where, in evaluating the leading WKB contribution,
the integration limits have been extended to
$0<x<\infty$. At fixed $\xi_{\rm eff}$ and $H_{\rm inf}$,
the magnetic integral scale is independent of
the overall visible-sector projection.
The corresponding energy-weighted magnetic
wavelength is
\begin{equation}
	\lambda_{B,\rm end}^{\rm phys}
	=2\pi L_{c,\rm end}^{\rm phys}.
	\label{eq:lambdaB_end}
\end{equation}
The corresponding comoving integral scale is
$L_{c,\rm end}=L_{c,\rm end}^{\rm phys}/a_{\rm end}$.
Throughout this work, we distinguish the magnetic
integral scale from the energy-weighted wavelength
$\lambda_B=2\pi L_c$. 

For the illustrative fixed-background benchmark
$\xi_{\rm eff}\simeq4.7$ and
$H_{\rm inf}\simeq5.88\times10^{-6}M_{\rm Pl}$
obtained in Sec.~\ref{sec:gauge_production}, using
$M_{\rm Pl}=2.435\times10^{18}\,{\rm GeV}$
and $1\,{\rm G}\simeq1.95\times10^{-20}\,{\rm GeV}^2$,
the leading WKB estimates give
\begin{align}
	B_{\rm end}
	&\simeq8.43\times10^2
	|g_{\rm eff}|H_{\rm inf}^2
	\nonumber\\
	&\simeq8.86\times10^{48}
	|g_{\rm eff}|\,{\rm G},
\end{align}
and
\begin{equation}
	L_{c,\rm end}^{\rm phys}
	\simeq\frac{2.69}{H_{\rm inf}}
	\simeq3.70\times10^{-29}\,{\rm m}.
\end{equation}
This benchmark corresponds to the nominal onset
of significant background backreaction.
For comparison, the conservative reference
$\xi_{\rm eff}\simeq3.87$ gives
$B_{\rm end}\simeq1.06\times10^{48}
|g_{\rm eff}|\,{\rm G}$ and
$L_{c,\rm end}^{\rm phys}
\simeq3.05\times10^{-29}\,{\rm m}$.

The constant-mixing analysis reveals a trade-off
between suppressing Schwinger dissipation and
efficient visible magnetic-field generation.
For a fixed visible magnetic amplitude, reducing
$|g_{\rm eff}|$ requires a larger canonical dark
magnetic energy, potentially enhancing background
backreaction. This motivates a phenomenologically
evolving kinetic portal that remains weak during
most of inflation but approaches
$|g_{\rm eff,end}|\sim\mathcal{O}(1)$ towards its
termination. Provided that the canonical visible
contribution remains subdominant, the endpoint
magnetic amplitude satisfies
$B_{\rm end}\simeq
|g_{\rm eff,end}|B_{\widetilde X,\rm end}$,
and hence becomes comparable to the canonical
dark magnetic amplitude. Such an endpoint does
not introduce a parametrically small gauge kinetic
eigenvalue or a nonperturbative effective charge.

A simple phenomenological realization is
\begin{equation}\label{eq:geff-pheno}
	g_{\rm eff}(N)=g_*e^{-pN},\qquad p>0, 
\end{equation}
Here, $N\equiv\ln(a_{\rm end}/a)$ denotes the
number of e-folds remaining until the end of
inflation, with $N=0$ at its termination and
$N\simeq60$ at the reference inflationary epoch.
The endpoint mixing is
$g_*\equiv g_{\rm eff,end}$, with
$|g_*|\sim\mathcal{O}(1)$.
The prescribed evolution corresponds to
$\dot g_{\rm eff}=pHg_{\rm eff}$.
Along a monotonically decreasing inflaton trajectory,
the corresponding scalar-dependent portal is
\begin{equation*}
	g_{\rm eff}(\phi)=g_*
	\exp\left[-\frac{p}{M_{\rm Pl}}
	\int_{\phi_{\rm end}}^\phi
	\frac{d\varphi}{\sqrt{2\epsilon_H(\varphi)}}\right].
\end{equation*}
This constitutes a phenomenological realization of
the scalar-dependent kinetic mixing in Eq.~(\ref{eq:action}).
From an effective-field-theory perspective, a transient
kinetic portal may arise through a gauge-singlet
auxiliary scalar $\chi$ coupled via the operator
$-c_\chi\chi F_{\mu\nu}Y^{\mu\nu}/(2\Lambda)$,
where $\Lambda$ denotes the effective suppression scale.
An inflaton-dependent minimum of $\chi$ may activate
the mixing towards the end of inflation, while its
subsequent relaxation suppresses the portal during
reheating. We adopt a phenomenological description
of this possibility without specifying its microscopic
completion. The endpoint projection remains an analytical
estimate, whose validity must be assessed through the
coupled gauge-field evolution in Sec.~\ref{sec:numerical}.

\section{Post-inflationary Evolution}
\label{sec:postinflation}

The inflationary kinetic-mixing prescription is
reconstructed along the monotonic scalar trajectory
and is not assumed to remain valid during a
subsequent oscillatory reheating phase.

To investigate the consequences of a transient
kinetic portal, we independently parametrize the
post-inflationary mixing during radiation domination as
\begin{equation}
	g_{\rm eff}(a)=g_{\rm eff,end}
	\left(\frac{a_{\rm end}}{a}\right)^q,
	\qquad q>0,
	\label{eq:postinflation_mixing}
\end{equation}
within the instantaneous-reheating reference
scenario.
The corresponding kinetic-mixing parameter,
$\epsilon_{\rm mix}=g_{\rm eff}/
\sqrt{1+g_{\rm eff}^2}$,
also approaches zero, rendering the visible and
dark gauge sectors effectively decoupled at late
times.

This parametrization is not intended to describe
the detailed inflaton oscillations or to provide
a microscopic completion of the mixing evolution
through reheating.

Writing $s\equiv\epsilon_{\rm mix}$, the transverse
visible-field equation in the presence of a
homogeneous physical conductivity
$\sigma_{\rm phys}$ becomes
\begin{equation}
	(A_\lambda'+sX_\lambda')'
	+k^2(A_\lambda+sX_\lambda)
	+a\sigma_{\rm phys}A_\lambda'=0.
\end{equation}
During radiation domination, neglecting changes in
the relativistic degrees of freedom, we approximate
$a\sigma_{\rm phys}\simeq\Gamma={\rm const.}$
Defining $\Pi_A=A_\lambda'+sX_\lambda'$, integration
of the visible-field equation gives
\begin{equation}
	\Gamma\Delta A_\lambda
	=
	-\Delta\Pi_A
	-k^2\int_{\tau_{\rm end}}^\tau
	(A_\lambda+sX_\lambda)\,d\tau'.
	\label{eq:magnetic_survival}
\end{equation}
Consequently, the previously generated visible magnetic
configuration approximately survives the decay of
kinetic mixing provided
\begin{equation}
	\frac{
		|\Delta\Pi_A|
		+k^2\displaystyle\int_{\tau_{\rm end}}^\tau
		|A_\lambda+sX_\lambda|\,d\tau'
	}{
		a_{\rm end}\sigma_{\rm end}|A_{\lambda,\rm end}|
	}
	\ll1.
\end{equation}
Equation~(58) provides a sufficient condition for
the survival of the comoving visible magnetic field
during the decay of the kinetic portal. Although
a highly conducting plasma with
$\sigma_{\rm phys}\gg H,k/a$ suppresses magnetic
diffusion, magnetic survival depends on the
integrated coupled-field evolution rather than
solely on the instantaneous kinetic mixing.
We assume that this condition is satisfied in
the following estimates \cite{Kamada:2018kyi}.

For illustrative estimates, we adopt instantaneous
reheating, independently of the microscopic reheating
mechanism. With
$\rho_{\rm end}\simeq3M_{\rm Pl}^2H_{\rm end}^2$
and neglecting the nonthermal energy fraction,
the corresponding reheating temperature and scale
factor are, respectively,
\begin{align}
	T_{\rm reh}
	&=
	\left(\frac{90}{\pi^2g_{\rm reh}}\right)^{1/4}
	\sqrt{M_{\rm Pl}H_{\rm end}},
	\label{eq:Treh}
	\\
	a_{\rm end}
	&=
	\frac{T_0}{T_{\rm reh}}
	\left(\frac{g_{s,0}}{g_{s,\rm reh}}\right)^{1/3},
	\label{eq:aend}
\end{align}
where $a_0=1$, and $g$ and $g_s$ denote the
effective relativistic degrees of freedom associated
with the energy and entropy densities, respectively.

Before nonlinear processing, the comoving magnetic
amplitude and integral scale,
$B_c=a^2B_{\rm phys}$ and
$L_c=L_c^{\rm phys}/a$, remain approximately
constant. Nonlinear magnetic processing becomes
effective when the Alfv\'en crossing time becomes
comparable to the Hubble time.
In the weak-field limit, the Alfv\'en velocity
during radiation domination is
\begin{equation}
	V_A\simeq
	\frac{B_{\rm end}}
	{\sqrt{\rho_{\rm end}+p_{\rm end}}}
	=
	\sqrt{\frac{3}{4}}
	\frac{B_{\rm end}}{\sqrt{\rho_{\rm end}}}.
	\label{eq:alfven_velocity}
\end{equation}
Here, $\rho_{\rm end}$ and $p_{\rm end}$ denote
the energy density and pressure of the radiation
plasma immediately after instantaneous reheating,
with $p_{\rm end}=\rho_{\rm end}/3$.
The magnetic amplitude $B_{\rm end}$ is obtained
in Sec.~\ref{sec:magnetic_end}, assuming its survival through the
transition to the conducting plasma.
Since $V_A$ remains approximately constant and
$HL_c^{\rm phys}\propto a^{-1}$ before nonlinear
processing during radiation domination, the
scale factor at its onset is approximately
\begin{equation}
	a_{\rm nl}\simeq r a_{\rm end},
	\qquad
	r\equiv
	\max\left(
	1,\frac{H_{\rm end}L_{c,\rm end}^{\rm phys}}{V_A}
	\right),
	\label{eq:nonlinear_onset}
\end{equation}
where changes in the relativistic degrees of
freedom before nonlinear processing are neglected.

For an approximately maximally helical magnetic
configuration, conservation of comoving magnetic
helicity implies $B_c^2L_c\simeq\mathrm{constant}$.
The standard helical turbulent-decay laws during
radiation domination are
\begin{equation}
	B_c\propto a^{-1/3},
	\qquad
	L_c\propto a^{2/3},
	\label{eq:helical_decay}
\end{equation}
as established in studies of decaying primordial
magnetic turbulence
\cite{Banerjee:2004df}.
We approximate
the termination of nonlinear processing by
matter--radiation equality, denoted by $a_m$.
For $a_{\rm nl}<a_m$, the present-day magnetic
amplitude and integral scale are therefore
\begin{align}
	&B_0
	\simeq
	a_{\rm end}^2 B_{\rm end}
	\left(\frac{a_m}{a_{\rm end}}\right)^{-1/3}
	r^{1/3},
	\label{eq:B_today}
	\\
	&L_{c,0}
	\simeq
	\frac{L_{c,\rm end}^{\rm phys}}{a_{\rm end}}
	\left(\frac{a_m}{a_{\rm end}}\right)^{2/3}
	r^{-2/3}.
	\label{eq:Lc_today}
\end{align}
If $a_{\rm nl}\geq a_m$, we instead retain
flux-freezing evolution within this prescription.

Adopting $T_0=2.7255\,{\rm K}$,
$g_{s,0}=g_{s,m}=3.94$,
$g_{\rm reh}=g_{s,\rm reh}=106.75$, and the
reference equality temperature $T_m=1\,{\rm eV}$, we obtain
\begin{align}
	B_0
	&\simeq
	1.175\times10^{-14}\,{\rm G}
	\frac{B_{\rm end}}{\sqrt{\rho_{\rm end}}}
	\left(\frac{10^{-6}M_{\rm Pl}}{H_{\rm end}}\right)^{1/6}
	r^{1/3},
	\label{eq:B_today_general}
	\\
	L_{c,0}
	&\simeq
	1.105\,{\rm pc}\,
	\left(H_{\rm end}L_{c,\rm end}^{\rm phys}\right)
	\left(\frac{10^{-6}M_{\rm Pl}}{H_{\rm end}}\right)^{1/6}
	r^{-2/3}.
	\label{eq:Lc_today_general}
\end{align}
The numerical coefficients follow from our
instantaneous-reheating prescription and the
adopted relativistic degrees of freedom, while
the delayed-onset scaling relations follow
Ref.~\cite{Sharma:2024nfu}.

For the illustrative analytical benchmark of Sec.~\ref{sec:magnetic_end},
we adopt $H_{\rm end}\simeq H_{\rm inf}$, with
$\xi_{\rm eff}\simeq4.70$ and
$H_{\rm inf}\simeq5.88\times10^{-6}M_{\rm Pl}$,
we obtain
\begin{align}
	T_{\rm reh}
	&\simeq3.19\times10^{15}\,{\rm GeV},
	\\
	a_{\rm end}
	&\simeq2.45\times10^{-29},
	\\
	V_A
	&\simeq2.48\times10^{-3}|g_{\rm eff,end}|,
	\\
	r
	&\simeq1.08\times10^{3}|g_{\rm eff,end}|^{-1}.
	\label{eq:r_benchmark}
\end{align}
Consequently, for delayed nonlinear processing
before equality, the present-day magnetic
amplitude and integral scale become
\begin{align}
	B_0
	&\simeq
	2.57\times10^{-16}
	|g_{\rm eff,end}|^{2/3}\,{\rm G},
	\label{eq:B_today_final}
	\\
	L_{c,0}
	&\simeq
	2.09\times10^{-2}
	|g_{\rm eff,end}|^{2/3}\,{\rm pc}.
	\label{eq:Lc_today_final}
\end{align}
The corresponding energy-weighted magnetic
wavelength is
\begin{equation}
	\lambda_{B,0}=2\pi L_{c,0}
	\simeq0.132|g_{\rm eff,end}|^{2/3}\,{\rm pc}.
	\label{eq:lambdaB_today}
\end{equation}
Thus, at a given canonical endpoint spectrum and
$H_{\rm inf}$, and provided that
$a_{\rm end}<a_{\rm nl}<a_m$,
both the present-day magnetic amplitude and
its integral scale exhibit the characteristic
scaling $|g_{\rm eff,end}|^{2/3}$.
This dependence originates from the inflationary
visible-sector projection and the associated delay
in nonlinear magnetic processing, rather than
from the residual late-time kinetic mixing. For $|g_{\rm eff,end}|\sim\mathcal{O}(1)$, these
estimates yield magnetic fields of order
$10^{-16}\,{\rm G}$ with subparsec wavelengths.

For comparison, the conservative analytical
benchmark $\xi_{\rm eff}\simeq3.87$ gives
$B_0\simeq5.86\times10^{-17}
|g_{\rm eff,end}|^{2/3}\,{\rm G}$
and $\lambda_{B,0}\simeq3.00\times10^{-2}
|g_{\rm eff,end}|^{2/3}\,{\rm pc}$.

The residual effective mixing at the onset of
nonlinear processing is
\begin{equation}
	g_{\rm eff,nl}=g_{\rm eff,end}r^{-q}.
\end{equation}
For sufficiently large $q$, the visible and dark
sectors become effectively decoupled before
nonlinear magnetic processing. Provided that
Eq.~(58) is satisfied, the subsequent magnetic
evolution becomes approximately insensitive
to the detailed decay of the kinetic portal.

To illustrate the sensitivity to a finite reheating
duration, consider an intervening matter-dominated
phase with no additional gauge-field amplification
or dissipation.
Defining $\vartheta=T_{\rm reh}/T_{\rm inst}$,
where $T_{\rm inst}$ denotes the instantaneous
reheating temperature, and neglecting changes in
the relativistic degrees of freedom, we obtain
\begin{equation}
	\frac{a_{\rm reh}}{a_{\rm end}}
	=\vartheta^{-4/3},
	\qquad
	\frac{a_{\rm end}}{a_{\rm end}^{\rm inst}}
	=\vartheta^{1/3}.
\end{equation}
Assuming adiabatic magnetic redshifting during
reheating, one has
$H_{\rm reh}L_{c,\rm reh}^{\rm phys}
\propto\vartheta^{2/3}$ and
$V_{A,\rm reh}\propto\vartheta^{2/3}$.
The nonlinear delay factor $r$ therefore remains
unchanged relative to the instantaneous reference,
with $a_{\rm nl}=ra_{\rm reh}$. For nonlinear
processing before matter--radiation equality,
we obtain
\begin{equation}
	\boxed{
		\frac{B_0}{B_0^{\rm inst}}
		=
		\frac{L_{c,0}}{L_{c,0}^{\rm inst}}
		=
		\left(\frac{T_{\rm reh}}{T_{\rm inst}}\right)^{1/3}.
	}
	\label{eq:reheating_scaling}
\end{equation}
For the conservative analytical benchmark considered
above, $T_{\rm reh}=10^{10}\,{\rm GeV}$ reduces the
magnetic estimates to
$B_0\simeq8.6\times10^{-19}
|g_{\rm eff,end}|^{2/3}\,{\rm G}$ and
$\lambda_{B,0}\simeq4.4\times10^{-4}
|g_{\rm eff,end}|^{2/3}\,{\rm pc}$.

\section{Numerical illustration: Starobinsky inflation}
\label{sec:numerical}

We illustrate the preceding framework using Starobinsky inflation with
exponential kinetic mixing, evolving the gauge fields on a scalar-only
background in the absence of charged currents.

\subsection{Inflationary background and kinetic mixing}

\begin{figure}[t]
	\centering
	\includegraphics[width=\columnwidth]
	{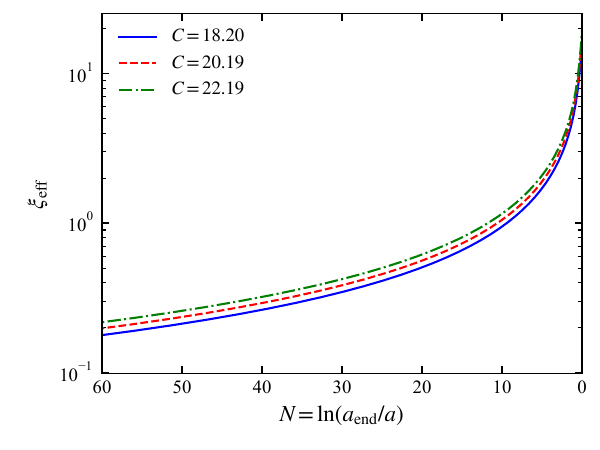}
	\caption{Evolution of the instantaneous instability
		parameter $\xi_{\rm eff}$ as a function of the
		number of e-folds remaining until the end of
		inflation for representative axial couplings.}
	\label{fig:xi_evolution}
\end{figure}

We consider the Starobinsky potential~\cite{Starobinsky:1980te},
\begin{equation}
	V(\phi)=\frac34 m^2M_{\rm Pl}^2
	\left(1-e^{-\sqrt{2/3}\phi/M_{\rm Pl}}\right)^2.
	\label{eq:starobinsky_potential}
\end{equation}
Normalizing the vacuum scalar spectrum
$\mathcal P_{\mathcal R}=H_*^2/(8\pi^2M_{\rm Pl}^2\epsilon_{H,*})$
to $A_s\simeq2.1\times10^{-9}$ at the reference epoch $N_*=60$,
we obtain
\begin{align}
	m &\simeq 1.1434\times10^{-5}M_{\rm Pl},\\
	H_* &\simeq 5.6488\times10^{-6}M_{\rm Pl},\\
	H_{\rm end} &\simeq 2.7628\times10^{-6}M_{\rm Pl}.
	\label{eq:numerical_background}
\end{align}
The choice $N_*=60$ serves as a representative
inflationary normalization epoch; its precise
relation to the observed CMB pivot depends on
the reheating history. We adopt $g_{\rm eff}=g_*e^{-pN}$ with $g_*=0.5$ and $p=1$,
where $N=\ln(a_{\rm end}/a)$. Defining $C=\alpha M_{\rm Pl}/f$,
the instantaneous axial instability parameter is
\begin{equation}
	|\xi_{\rm eff}|=|C|(1+g_{\rm eff}^2)
	\sqrt{\frac{\epsilon_H}{2}}.
\end{equation}
Its evolution is shown in Fig.~\ref{fig:xi_evolution}. At the reference CMB epoch $N_*=60$, the instability
parameter remains small, $|\xi_{\rm eff,*}|\sim0.2$,
for the representative axial couplings.
Gauge-sourced scalar and tensor contributions around
CMB horizon exit are therefore expected to be
negligible \cite{Barnaby:2010vf, Barnaby:2011qe}.

\subsection{Gauge-field evolution and backreaction}

\begin{figure}[t]
	\centering
	\includegraphics[width=\columnwidth]
	{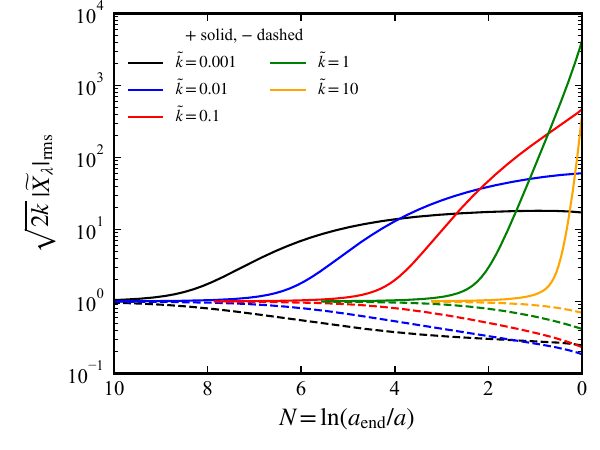}
\caption{Evolution of the canonical dark gauge-field
	mode amplitudes
	$\sqrt{2k\sum_j|\widetilde X_{\lambda j}|^2}$
	for representative dimensionless comoving momenta
	$\tilde{k}=k/(a_{\rm end}H_{\rm end})$ at $C=18.20$.
	Solid and dashed curves correspond to positive and
	negative helicities, respectively.}
	\label{fig:dark_modes}
\end{figure}

\begin{figure}[t]
	\centering
	\includegraphics[width=\columnwidth]
	{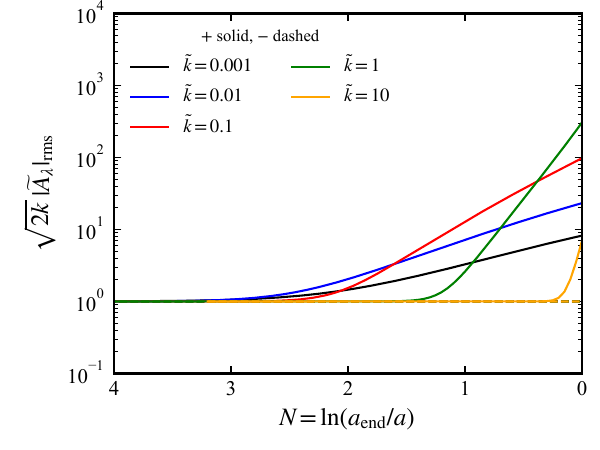}
	\caption{Evolution of the canonical visible gauge-field
		mode amplitudes
		$\sqrt{2k\sum_j|\widetilde A_{\lambda j}|^2}$.
		The parameters and line conventions are the same
		as in Fig.~\ref{fig:dark_modes}.}
	\label{fig:visible_modes}
\end{figure}

\begin{figure}[t]
	\centering
	\includegraphics[width=\columnwidth]
	{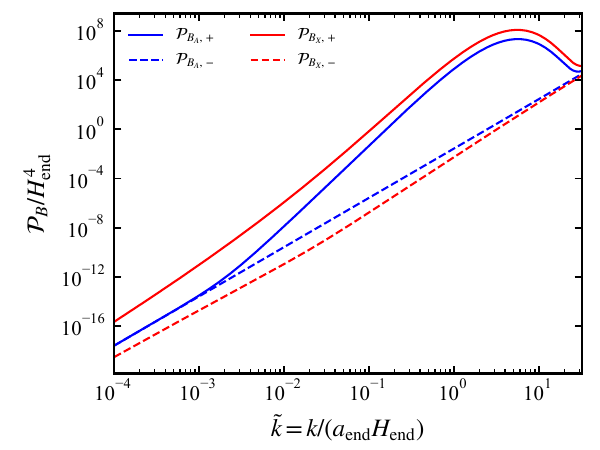}
	\caption{Magnetic power spectra of the visible and
		dark gauge fields in the interaction basis at
		the end of inflation for $C=18.20$.
		Both helicities are displayed separately.
		The spectra are normalized by $H_{\rm end}^4$.}
	\label{fig:magnetic_spectra}
\end{figure}

For evolving mixing, the canonical fields are dynamically coupled.
Following the coupled-field quantization of Ref.~\cite{Ferreira:2014mwa},
we expand the interaction-basis modes as
\begin{equation}
	\begin{pmatrix}\widehat A_\lambda\\\widehat X_\lambda\end{pmatrix}
	=\frac{U_\lambda}{\sqrt{2k}}
	\begin{pmatrix}\hat a_{\lambda1}\\\hat a_{\lambda2}\end{pmatrix}
	+\mathrm{h.c.},
\end{equation}
where the columns of $U_\lambda$ describe two independent initial
quantum oscillators and spatial Fourier factors are implicit.
Writing $s=\epsilon_{\rm mix}$, we define
\begin{equation}
	K=\begin{pmatrix}1&s\\s&1\end{pmatrix},\quad
	D=\begin{pmatrix}0&0\\0&1\end{pmatrix},\quad
	P_\lambda=KU_\lambda'/k.
\end{equation}
The coupled evolution is governed by
\begin{align}
	\frac{dU_\lambda}{dN}&=-xK^{-1}P_\lambda,\nonumber\\
	\frac{dP_\lambda}{dN}&=xKU_\lambda+\lambda CvDU_\lambda,
	\label{eq:numerical_firstorder}
\end{align}
where $x=k/(aH)$ and $v=\dot\phi/(HM_{\rm Pl})$. The columns of $U_\lambda$ are evolved independently
under the same coupled matrix equations, preserving
the canonical normalization of the two initial
quantum oscillators.
We impose the leading adiabatic vacuum deep inside the horizon,
with $U_\lambda^\dagger P_\lambda-P_\lambda^\dagger U_\lambda=-2iI$
and $U_\lambda^TP_\lambda-P_\lambda^TU_\lambda=0$. We initialize each mode at $x_i=140$ using
the leading instantaneous adiabatic conditions
\begin{align*}
	U_{\lambda,i}
	&=T_i\,{\rm diag}(1,w_{\lambda,i}^{-1/2}),\\
	P_{\lambda,i}
	&=-iK_iT_i\,{\rm diag}(1,w_{\lambda,i}^{1/2}),
\end{align*}
where
\begin{equation*}
	T=
	\begin{pmatrix}
		1 & -s/\beta\\
		0 & 1/\beta
	\end{pmatrix},
	\qquad
	w_{\lambda,i}=
	\sqrt{1-\frac{2\lambda\xi_{\rm eff,i}}{x_i}}.
\end{equation*}
The instantaneous canonical modes follow from Eq.~(\ref{eq:canonical_transform}).
Their evolution and the magnetic spectra are shown in
Figs.~\ref{fig:dark_modes}--\ref{fig:magnetic_spectra}.


\begin{figure}[t]
	\centering
	\includegraphics[width=\columnwidth]
	{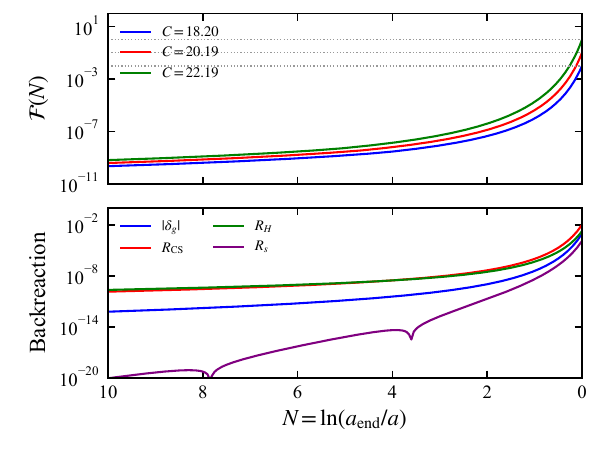}
	\caption{Evolution of the background-backreaction
		diagnostics during inflation. The upper panel shows
		the total diagnostic $\mathcal{F}(N)$ for the three
		representative axial couplings $C=18.20$, $20.19$,
		and $22.19$. The lower panel displays the gauge-energy
		fraction $|\delta_g|$, the Chern--Simons source
		$R_{\rm CS}$, the Hubble backreaction ratio $R_H$,
		and the mixing-induced source $R_s$ for $C=18.20$.}
	\label{fig:backreaction}
\end{figure}

The gauge energy density is
\begin{align}
	\rho_g={}&\frac12\left\langle
	E_A^2+B_A^2+E_X^2+B_X^2\right\rangle\nonumber\\
	&+s\left\langle\mathbf E_A\cdot\mathbf E_X+
	\mathbf B_A\cdot\mathbf B_X\right\rangle.
	\label{eq:numerical_energy}
\end{align}
For the numerical estimates, we subtract the
leading free-vacuum contribution within the
instantaneous tachyonic band
$0<x<2|\xi_{\rm eff}|$, rather than performing
a full adiabatic renormalization. This differs
from the leading exponentially enhanced
approximation adopted in Sec.~\ref{sec:gauge_production}.
The finite-band prescription does not account
for possible gauge-field production outside
this band.
With $\varphi=\phi/M_{\rm Pl}$, the backreaction diagnostics are
\begin{align}
	\delta_g&=\frac{\rho_g}{3M_{\rm Pl}^2H^2},\qquad
	R_H=\frac{2|\delta_g|}{\epsilon_H},\nonumber\\
	R_{\rm CS}&=\frac{|C\langle\mathbf E_X\cdot\mathbf B_X\rangle|}
	{3M_{\rm Pl}^2H^2|v|},\nonumber\\
	R_s&=\frac{|s_{,\varphi}\langle\mathbf E_A\cdot\mathbf E_X-
		\mathbf B_A\cdot\mathbf B_X\rangle|}{3M_{\rm Pl}^2H^2|v|},
\end{align}
where $s_{,\phi}=pg_{\rm eff}/
[v(1+g_{\rm eff}^2)^{3/2}]$ and all background
quantities are evaluated on the scalar-only
solution.
Since $R_{\rm CS}+R_s$ provides a conservative
estimate of the total scalar backreaction without
relying on cancellations between the two sources,
we define the maximum background-backreaction
diagnostic as
\begin{equation}
	\begin{aligned}
		\mathcal F(C)&=\max_{0\leq N\leq60}\mathcal F(N;C),\\
		\mathcal F(N;C)&=\max\{|\delta_g|,R_{\rm CS}+R_s,R_H\}.
	\end{aligned}
\end{equation}
Figure \ref{fig:backreaction} illustrates the resulting backreaction
diagnostics. Adopting $\mathcal F\simeq1$ as the
nominal onset of significant background backreaction,
we obtain
\begin{equation}
	\left.\frac{|\alpha|}{f}\right|_{\rm BR}
	\simeq\frac{22.19}{M_{\rm Pl}}.
\end{equation}
For comparison, $\mathcal F\simeq0.01$ and $0.1$
correspond to $|C|\simeq18.20$ and $20.19$,
respectively. The mixing-induced scalar source
constitutes approximately one percent of the axial
source at the endpoint for these benchmarks.
These estimates diagnose the onset of background
backreaction rather than perturbative consistency.
In particular, the instability parameter reaches
$|\xi_{\rm eff,end}|\simeq16.09$ for $|C|\simeq18.20$.
Such a rapid enhancement towards the end of inflation
may generate appreciable small-scale curvature
perturbations. The conventional constant-$\xi_{\rm eff}$
inverse-decay estimates cannot be directly applied
to the rapidly evolving coupled gauge system
considered here. A dedicated investigation of
sourced perturbations, their non-Gaussian statistics
and the associated primordial black hole abundance
is deferred to future work.

\subsection{Magnetic-field estimates}

The visible magnetic spectrum is obtained directly from the
interaction-basis modes,
\begin{equation}
	\mathcal P_{B_A}(k,\tau)=\frac{k^4}{4\pi^2a^4}
	\sum_{\lambda,j}|U_{A,\lambda j}|^2,
\end{equation}
and integrated over the adopted production band using the definitions
of Sec.~\ref{sec:magnetic_end}. For the reference benchmark $g_*=0.5$, $p=1$
and $|C|\simeq18.20$,
$B_{\rm end}\simeq1.14\times10^{49}\,\mathrm G$ and
$L_{c,\rm end}^{\rm phys}\simeq6.05\times10^{-30}\,\mathrm m$.
Using Sec.~\ref{sec:postinflation}, instantaneous reheating gives
$T_{\rm reh}\simeq2.19\times10^{15}\,\mathrm{GeV}$ and
$a_{\rm end}\simeq3.57\times10^{-29}$.
The resulting present-day estimates are listed in
Table~\ref{tab:magnetic_results}.

Although the instantaneous instability parameter
becomes substantially larger than the constant
$\xi_{\rm eff}$ reference towards the end of
Starobinsky inflation, the accumulated gauge-field
amplification depends on its full evolution rather
than its endpoint value. The short duration of
the large-instability regime, together with
$H_{\rm end}\simeq0.49H_*$, helps explain the
comparable orders of magnitude of the analytical
and numerical magnetic estimates.

\begin{table*}[t]
	\centering
	\renewcommand{\arraystretch}{1.15}
	\begin{tabular}{ccccccccc}
		\hline\hline
		$g_*$ & $p$ & $\mathcal F$ & $|C|$ &
		$B_0^{\rm inst}$ (G) &
		$\lambda_{B,0}^{\rm inst}$ (pc) &
		$B_0^{(10)}$ (G) &
		$\mathcal I_{\rm Sch,stat}^{(A)}$ &
		$\mathcal I_{\rm Sch,rel}^{(A)}$ \\
		\hline
		$0.1$ & $2$ & $0.01$ & $19.64$ &
		$8.50\times10^{-17}$ & $0.044$ &
		$1.41\times10^{-18}$ & $0.189$ & $0.028$ \\
		
		$0.25$ & $2$ & $0.01$ & $19.45$ &
		$1.53\times10^{-16}$ & $0.078$ &
		$2.54\times10^{-18}$ & $0.479$ & $0.069$ \\
		\hline
		$0.5$ & $1$ & $0.01$ & $18.20$ &
		$2.42\times10^{-16}$ & $0.124$ &
		$4.02\times10^{-18}$ & $1.089$ & $0.159$ \\
		
		$0.5$ & $1$ & $0.1$ & $20.19$ &
		$4.93\times10^{-16}$ & $0.253$ &
		$8.19\times10^{-18}$ & $3.166$ & $0.435$ \\
		
		$0.5$ & $1$ & $1$ & $22.19$ &
		$1.01\times10^{-15}$ & $0.516$ &
		$1.67\times10^{-17}$ & $9.283$ & $1.210$ \\
		\hline\hline
	\end{tabular}
	\caption{Current-free magnetic estimates and cumulative
		visible-normalized Schwinger diagnostics for representative
		kinetic-mixing histories. The superscript ``inst'' denotes
		instantaneous reheating, while $B_0^{(10)}$ corresponds to
		an intervening matter-dominated reheating phase with
		$T_{\rm reh}=10^{10}\,{\rm GeV}$.
		The subscripts ``stat'' and ``rel'' denote stationary
		conductivity and the phenomenological relaxation model,
		respectively. The first two benchmarks remain below the
		nominal background and cumulative dissipation thresholds,
		whereas the remaining benchmarks exceed the nominal
		visible-normalized stationary dissipation threshold.
		The last row represents the nominal onset of background
		backreaction. All magnetic amplitudes are indicative
		estimates without current-induced feedback.}
	\label{tab:magnetic_results}
\end{table*}

To estimate the cumulative importance of Schwinger
dissipation, we define
\begin{equation}
	\mathcal I_{\rm Sch}^{(i)}
	=\frac{1}{\rho_{i,\rm end}}
	\int_0^{N_i}dN\,e^{-4N}
	\frac{\sigma_{\rm Sch}E_A^2}{H},
	\qquad i=g,A,
\end{equation}
Here, $N_i$ denotes the initial e-fold value
of the dissipation integral, and $\rho_A$ denotes
the current-aligned visible energy density
defined in Sec.~\ref{sec:implications}. All quantities in the
integrand are evaluated on the current-free
gauge-field solution.
We adopt $\mathcal I_{\rm Sch}^{(i)}\sim1$ as the
nominal threshold for appreciable cumulative
dissipation relative to the corresponding energy
density.

For $|C|\simeq18.20$, the stationary conductivity
gives $\mathcal I_{\rm Sch}^{(g)}\simeq0.218$
and $\mathcal I_{\rm Sch}^{(A)}\simeq1.089$.
An illustrative phenomenological relaxation model,
$\dot j+3Hj=3H\sigma_{\rm inst}E_A$, with
$\sigma_{\rm Sch}=j/E_A$ and $\sigma_{\rm inst}$
given by Eq.~(\ref{eq:schwinger_conductivity}), reduces these estimates to
$0.032$ and $0.159$, respectively.  The relaxation current is initialized at zero
at $N_i=60$.
The strong-field criterion is imposed on its
source term. The relaxation rate $3H$ is motivated by the
cosmological dilution of the physical number density
of produced charge carriers. In the limit of rapidly
accelerated ultrarelativistic fermions, the conduction
current approximately follows this density.
Matching its stationary value to Eq.~(\ref{eq:schwinger_conductivity}) motivates
the relaxation prescription, which neglects the
detailed momentum evolution and polarization current.
In evaluating the stationary conductivity, we retain
its contribution only within the adopted strong-field
regime $g'E_A/H^2\geq1$, neglecting the weak-field
contribution in this diagnostic estimate.

An order-unity diagnostic indicates appreciable
dissipation rather than necessarily complete magnetic
screening. Quantifying the actual magnetic suppression
requires the self-consistent evolution of the induced
current and the coupled gauge fields.
For the post-inflationary decay exponent, $q=4$ gives residual mixing
below $10^{-2}$ at nonlinear onset for all three $g_*=0.5$,
$p=1$ benchmarks; magnetic
survival remains conditional on the criterion of Sec.~\ref{sec:postinflation}.

\begin{figure*}[t]
	\centering
	\includegraphics[width=0.85\textwidth]
	{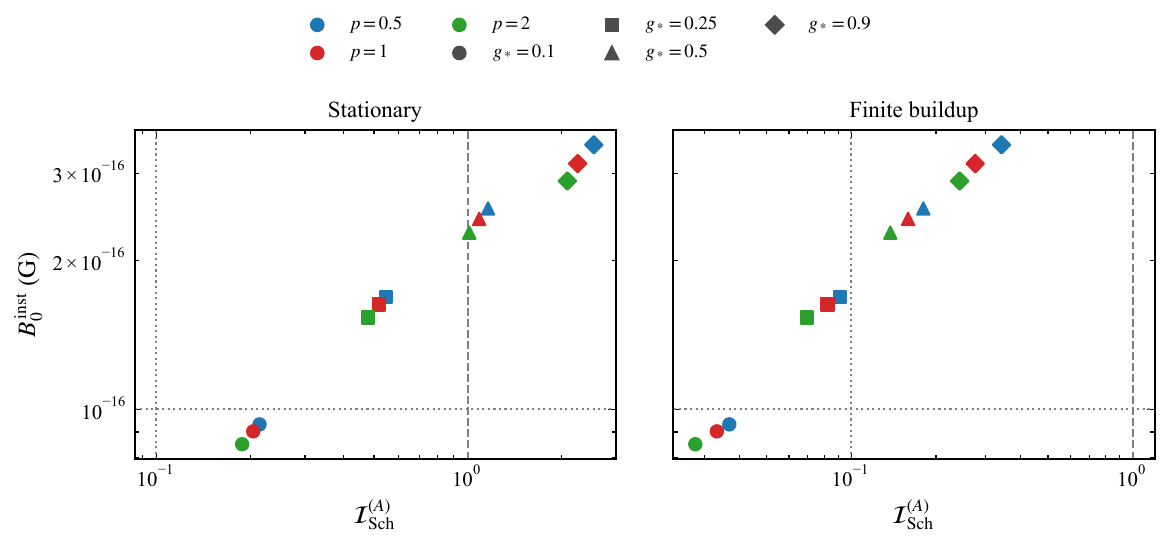}
	\caption{Present-day magnetic amplitudes as functions
		of the visible-normalized cumulative Schwinger
		diagnostic for stationary conductivity (left)
		and the phenomenological relaxation prescription
		(right). Each realization is calibrated to
		$\mathcal F\simeq0.01$. Different colors and markers
		represent the growth exponent $p$ and endpoint
		mixing $g_*$, respectively.
		The magnetic estimates assume instantaneous
		reheating and subsequent helical MHD evolution,
		without current-induced feedback.}
	\label{fig:portal_scan}
\end{figure*}

To illustrate the dependence on the kinetic portal,
we vary $g_*\in\{0.1,0.25,0.5,0.9\}$ and
$p\in\{0.5,1,2\}$ at a common background-backreaction
diagnostic $\mathcal F\simeq0.01$.
The results are shown in Fig.~\ref{fig:portal_scan}.

For $g_*=0.1$ and $p=2$, we obtain
$B_0\simeq8.50\times10^{-17}\,{\rm G}$ with
$\mathcal I_{\rm Sch}^{(g)}\simeq0.002$ and
$\mathcal I_{\rm Sch}^{(A)}\simeq0.189$
for stationary conductivity.
Increasing the endpoint mixing to $g_*=0.25$
yields $B_0\simeq1.53\times10^{-16}\,{\rm G}$,
with the corresponding diagnostics increasing to
$0.028$ and $0.479$, respectively.
Both realizations remain below the nominal
cumulative dissipation threshold, illustrating
the additional freedom provided by the kinetic
portal in balancing magnetic production against
charged-particle dissipation.

Within the parameter range investigated, varying the
kinetic portal changes the present-day magnetic
amplitude by approximately a factor of four, whereas
reducing the reheating temperature from its
instantaneous value to $10^{10}\,{\rm GeV}$ suppresses
the amplitude by approximately a factor of sixty.
The reheating history therefore has a stronger impact
on the attainable magnetic amplitude over the
considered range, while the kinetic portal primarily
controls the trade-off between magnetic production
and charged-particle dissipation.

The framework thus illustrates the potential for
helical fields of order $10^{-16}\,{\rm G}$ at
subparsec wavelengths for representative mixing
histories below the nominal background and
cumulative dissipation thresholds.
Current-free estimates approaching $10^{-15}\,{\rm G}$
occur near the onset of background backreaction and
exceed the nominal stationary dissipation threshold.
These indicative magnetic estimates cannot be
compared directly with megaparsec-scale magnetic
bounds
\cite{Caprini:2015gga,MAGIC:2022piy,HESS:2023zwb}.

\section{Astrophysical and cosmological implications}
\label{sec:implications}

The current-free magnetic estimates obtained in
Sec.~\ref{sec:numerical} reach amplitudes of order $10^{-16}\,{\rm G}$
with subparsec wavelengths for representative
kinetic portals. Such helical configurations may
provide primordial seeds for subsequent
astrophysical amplification, although direct
comparison with intergalactic magnetic-field
constraints requires accounting for their
correlation scale. Beyond primordial magnetic
seeds, the same inflationary gauge-field
configuration provides potential connections
to dark radiation and helicity-dependent
cosmological phenomena.

\paragraph{Dark radiation.}
The massless dark gauge field can contribute to the
effective number of relativistic species.
Completing the square in the gauge-sector energy density,
we obtain the instantaneous partition
\begin{align}
	\rho_{\mathcal A}
	&=\frac{\beta^2}{2}
	\left\langle
	\mathbf E_A^2+\mathbf B_A^2
	\right\rangle,
	\\
	\rho_{\mathcal X}
	&=\frac12\left\langle
	(\mathbf E_X+s\mathbf E_A)^2+
	(\mathbf B_X+s\mathbf B_A)^2
	\right\rangle,
\end{align}
such that $\rho_g=\rho_A+\rho_X$, with the
expectation values evaluated using the
vacuum prescription of Sec.~\ref{sec:numerical}.
For dominant canonical dark-field amplification,
the constant-mixing approximation gives
$\rho_A/\rho_g\simeq g_{\rm eff}^2/(1+g_{\rm eff}^2)$.
For $g_*=0.5$, this yields an endpoint visible energy
fraction of approximately $0.2$, in agreement with
the numerical results. This instantaneous energy
partition need not coincide with the final partition
after kinetic mixing becomes negligible.

Assuming instantaneous reheating and negligible
subsequent energy exchange between the surviving dark
radiation and the thermal plasma, its contribution
to the effective number of relativistic species is
\begin{equation}
	\Delta N_{\rm eff}\simeq
	\frac{4}{7}g_{*,{\rm reh}}
	\left(\frac{10.75}{g_{*s,{\rm reh}}}\right)^{4/3}
	\frac{\delta_{\mathcal X}}{1-\delta_{\mathcal X}}
	\simeq
	2.86\frac{\delta_{\mathcal X}}{1-\delta_{\mathcal X}},
\end{equation}
where $\delta_{\mathcal X}$ denotes the dark radiation
fraction of the total energy density after reheating.
Assuming that the orthogonal dark component survives
the reheating transition without significant energy
exchange, we approximate
\begin{equation}
	\delta_{\mathcal X}\simeq
	\frac{\rho_{\mathcal X,{\rm end}}}
	{3M_{\rm Pl}^2H_{\rm end}^2}.
\end{equation}
For the reference benchmark $g_*=0.5$, $p=1$
and $|C|\simeq18.20$,
$R_H\lesssim0.01$ and $\epsilon_{H,\rm end}=1$
imply $\delta_{g,\rm end}\lesssim5\times10^{-3}$.
Assuming negligible subsequent energy transfer and
$\delta_X\lesssim\delta_{g,\rm end}$, we obtain
\begin{equation}
	\Delta N_{\rm eff}\lesssim1.44\times10^{-2}.
\end{equation}
Under the adopted reheating and dark-sector decoupling
assumptions, the estimated contribution lies comfortably
within current BBN and CMB constraints on additional
relativistic species \cite{ParticleDataGroup:2024cfk, AtacamaCosmologyTelescope:2025nti}.

\paragraph{Magnetic helicity and baryogenesis.}
We quantify the helicity fraction of the visible
magnetic field at the end of inflation over the
adopted production band by
\begin{equation}
	\epsilon_h=
	\frac{
		\displaystyle\int d\ln k\,k^{-1}
		(\mathcal P_{B_A,+}-\mathcal P_{B_A,-})
	}{
		\displaystyle\int d\ln k\,k^{-1}
		(\mathcal P_{B_A,+}+\mathcal P_{B_A,-})
	}.
\end{equation}
The dominance of the positive-helicity magnetic
spectrum in Fig.~4 motivates the approximately
maximally helical configuration adopted for the
post-inflationary evolution. Under conservation of
comoving magnetic helicity, its magnitude is
\begin{equation}
	|h_c|\simeq|\epsilon_h|B_0^2L_{c,0}
	\sim10^{-33}-10^{-31}\,{\rm G}^2\,{\rm pc},
\end{equation}
for $|\epsilon_h|\simeq1$ and the current-free
$g_*=0.5$, $p=1$ benchmarks listed in Table~\ref{tab:magnetic_results}.

If the visible Abelian field is identified with
hypercharge above the electroweak scale, the
surviving hypermagnetic helicity can source
baryon number through the electroweak anomaly
\cite{Kamada:2016eeb,Domcke:2019mnd,Hamada:2025ooy}.
The conversion to electromagnetic fields,
chiral plasma effects and helicity redistribution
can modify the subsequent magnetic configuration.
We therefore identify the helical initial state
as a possible ingredient for baryogenesis, without
predicting the resulting baryon asymmetry or
electroweak transfer efficiency.

\paragraph{Chiral gravitational waves.}
The parity-violating anisotropic stresses of the
amplified gauge fields can also source chiral
gravitational waves.
For the reference benchmark $g_*=0.5$, $p=1$
and $|C|\simeq18.20$, the redshifted frequency
associated with the magnetic integral scale is
\begin{equation}
	f_{\rm src,0}\sim
	\frac{a_{\rm end}}
	{2\pi L_{c,\rm end}^{\rm phys}}
	\simeq3\times10^8\,{\rm Hz}.
\end{equation}
This characterizes the gauge-field source scale,
rather than a predicted gravitational-wave
spectral peak \cite{Sorbo:2011rz,Adshead:2019igv}.

\section{Discussion and conclusions}
\label{sec:conclusions}

We have developed an analytical framework for
inflationary helical magnetogenesis through a
kinetic portal connecting visible and dark
Abelian gauge fields.
Unlike conventional visible-sector axial
magnetogenesis, the amplification mechanism
operates in a dark gauge sector, while kinetic
mixing controls its interaction with charged
matter.

Within the constant-mixing approximation and
at a given canonical dark-field spectrum, the
visible magnetic power scales as $g_{\rm eff}^2$,
whereas the stationary estimate of Schwinger-induced
dissipation scales as $|g_{\rm eff}|^3$.
This parametric separation provides additional
freedom in controlling charged-particle
interactions.
Furthermore, the bounded kinetic eigenvalues
avoid the conventional strong-coupling pathology
associated with a vanishing gauge kinetic
normalization.
For delayed nonlinear magnetic processing before
matter--radiation equality, the adopted helical
MHD prescription yields the characteristic
dependence
$B_0,L_{c,0}\propto|g_{\rm eff,end}|^{2/3}$
at a given canonical endpoint spectrum.

We illustrated the analytical framework using the
phenomenological prescription $g_{\rm eff}=g_*e^{-pN}$
in a Starobinsky inflationary background.
For $g_*=0.25$ and $p=2$, the current-free calculation
yields $B_0\simeq1.53\times10^{-16}\,{\rm G}$ and
$\lambda_{B,0}\simeq0.078\,{\rm pc}$ for instantaneous
reheating, with both the homogeneous-background
and cumulative dissipation diagnostics below their
nominal thresholds. Larger magnetic estimates
approaching $10^{-15}\,{\rm G}$ exceed the nominal
stationary dissipation threshold.

The parameter scan demonstrates the additional
freedom provided by the kinetic portal in balancing
magnetic production against charged-particle
dissipation. The reheating history has a stronger
impact on the attainable magnetic amplitude over
the considered range. In particular, for an
intervening matter-dominated reheating phase
with adiabatic magnetic redshifting, we obtain
the conditional scaling
$B_0,L_{c,0}\propto T_{\rm reh}^{1/3}$.

The resulting helical gauge configurations also
provide potential connections to relic dark
radiation, baryogenesis and chiral gravitational
waves.
Their quantitative abundances depend on the
charged-particle response, the reheating
transition and subsequent plasma dynamics.
A quantitative determination of the surviving
magnetic amplitudes, sourced curvature perturbations
and the associated primordial black hole abundance
requires a dedicated investigation beyond the
present framework.

Our analytical estimates and numerical illustrations
demonstrate the potential of the kinetic portal to
alleviate, rather than eliminate, Schwinger dissipation.
This motivates further investigations of kinetically
mixed helical gauge sectors and their cosmological
consequences.

\section*{Acknowledgment}
The author is grateful to Ramkishor Sharma and
Nur Jaman for insightful and extremely valuable
discussions on various aspects of this work.

\end{document}